\documentclass[11pt,journal,onecolumn]{IEEEtran}
\usepackage{amssymb,cite}
\usepackage{amsmath,multirow,threeparttable,cite,graphicx,times,subfigure,url,verbatim, color, multirow}

\definecolor{red}{rgb}{0.8,0,0}

\begin{document}

\title{D2D Big Data: Content Deliveries over\\
Wireless Device-to-Device Sharing in \\Large Scale Mobile Networks}

\author{
    Xiaofei Wang$^{1*}$,
    Yuhua Zhang$^{1}$,
    Victor C. M. Leung$^{2}$,
    Nadra Guizani$^{3}$,
    Tianpeng Jiang$^{4}$\\
    \scriptsize
    ${^1}$Tianjin Key Laboratory of Advanced Networking (TANK), School of Computer Science and Technology,
    Tianjin University, Tianjin, China.\\
    ${^2}$Department of Computer Engineering, University of British Columbia, Vancouver, Canada.\\
    ${^3}$Department of Electrical and Computer Engineering, Purdue University, USA.\\
    ${^4}$Beijing Anqi Zhilian Technology Co. Ltd., Beijing, China.\\
    *Prof. Xiaofei Wang is the corresponding author.
}

\markboth{IEEE WIRELESS COMMUNICATIONS MAGAZINE, VOL. XX, NO. YY, MONTH XXXX}{}
\maketitle

\begin{abstract}

Recently the topic of \textit{how to effectively offload cellular traffic
onto device-to-device (D2D) sharing among users in proximity}
has been gaining more and more attention of global researchers and engineers.
Users utilize wireless short-range D2D communications for sharing contents locally,
due to not only the rapid sharing experience and free cost,
but also high accuracy on deliveries of interesting and popular contents,
as well as strong social impacts among friends.
Nevertheless, the existing related studies are
mostly confined to small-scale datasets, limited dimensions of user features,
or unrealistic assumptions and hypotheses on user behaviors.
In this article, driven by emerging Big Data techniques,
we propose to design a big data platform, named \textbf{D2D Big Data},
in order to encourage the wireless D2D communications among users effectively,
to promote contents for providers accurately,
and to carry out offloading intelligence for operators efficiently.
We deploy a big data platform and further utilize a large-scale dataset (3.56 TBytes)
from a popular D2D sharing application (APP),
which contains 866 million D2D sharing activities
on 4.5 million files disseminated via nearly 850 million users in 13 weeks.
By abstracting and analyzing multi-dimensional features,
including online behaviors, content properties, location relations,
structural characteristics, meeting dynamics, social arborescence, privacy preservation policies and so on,
we verify and evaluate the D2D Big Data platform regarding
predictive content propagating coverage.
Finally, we discuss challenges and opportunities regarding D2D Big Data
and propose to unveil a promising upcoming future of wireless D2D communications.

\end{abstract}


\section{Introduction}
\label{sec:intro}

With the increasing quantity and quality of multimedia services over mobile networks,
there has been a huge increase in traffic over recent years \cite{cisco_index_mobile_datatraffic}.
This has caused major hardships in effectively handling such traffic.
This is mainly due to the lack of scalable mobile network operators (MNOs)' infrastructure and the congested wireless network.

A recent new trend is to encourage users to obtain interesting
content files from the devices of other users around them.
This can be carried out by caching and sharing files
with the help of advanced device-to-device (D2D) communications,
which is an effective method to largely offload the duplicate cellular downloads
 \cite{d2djsac}.
D2D can thus significantly break the ``bottle-neck'' of the
MNOs' infrastructured networks.
Note that in this article, we consider various types of wireless D2D communication
techniques between devices directly,
even including the 3GPP D2D communication underlaying LTE networks.

Many studies, like 
\cite{xu_exploit_mobile} \cite{tagassisted_wirelesscommag},
have shown that by mining the social and mobile behaviors of users,
they are inclined to share popular content files
via frequently offline encounters over D2D communications \cite{TOSS_infocom14}.
However, all of the previous studies in the literature have focused
only on one particular group of users (in the scale of hundreds or thousands),
based on small-scale data analysis and algorithms design,
which cannot exploit D2D for main-stream
high-quality services rather than just a tool.
The promising D2D framework
for now should be significantly enhanced to be able to
elaborate the huge potential of D2D deliveries
over a large number of mobile users (in the scale of millions),
in a wide area (cities or even countries),
which motivates us to integrate state-of-the-art big data techniques
to encourage smart D2D deliveries among users effectively,
to promote contents for providers accurately,
and to carry out offloading intelligence for operators efficiently. This is
because there has been increasingly a vast amount of different types of data
(e.g., activity logs, traffic logs, profiles of users and contents, etc.)
not yet discovered and fully utilized.

\begin{figure}[!hhhhhhhhhht]
\centering
\includegraphics[width=12cm]{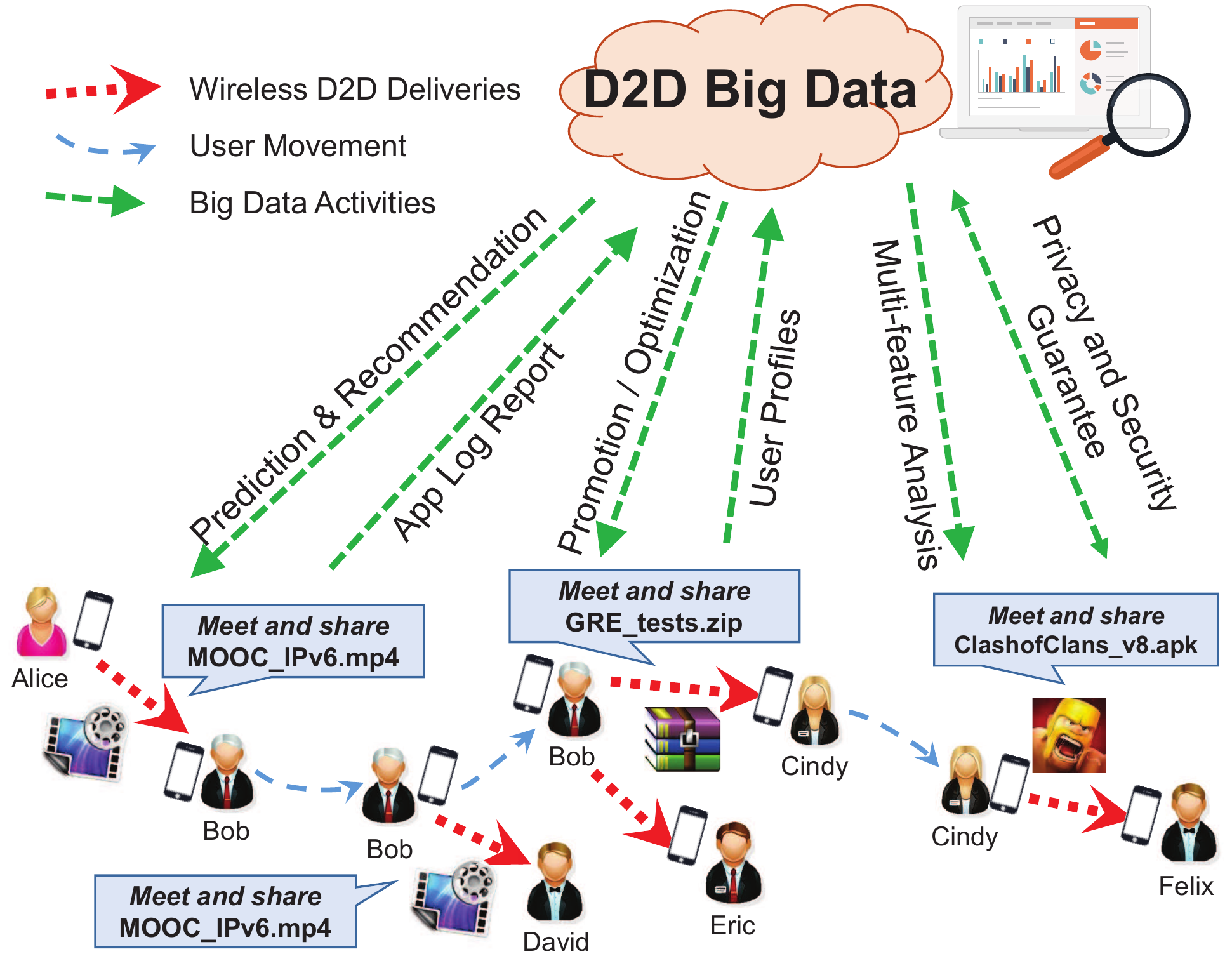}
\caption{Illustration of wireless D2D deliveries
integrated and supported with D2D Big Data}
\label{fig_1_offlinemsn.eps}
\end{figure}

The emerging big data techniques pose many new unprecedented
opportunities and challenges to conventional
data analytics over groups of mobile users for exploiting the capacity of D2D sharings
because of the significant dimensionality, heterogeneity, and complex features therein, e.g.,
that is, \emph{Volume, Variety, Velocity, Value and Veracity}
\cite{bigdata_classical_one}
\cite{yanzhang_socialmobilebigdata}.
Hence it is crucial to shift our attention from
single group-based research to a big data-based framework for D2D,
as illustrated in Fig. \ref{fig_1_offlinemsn.eps}.
First, the D2D platform can easily support
to adopt general statistical measurement algorithms
on the large scale user base (e.g., millions of people)
in order to grasp a common
understanding over the analysis results.
In fact, the utilization of \textbf{D2D Big Data} highly
depends on the exploration of multidimensional features extracted
from the complex mobile user behaviors,
e.g., geographical homophily of clustered users, social relationships,
interest similarity and mobility models,
and the complicated content properties, e.g., content size, categories,
and time-varying popularity.
Particularly, the sharing activities among mobile users can
formulate a connected graph so that D2D Big Data framework
can apply advanced theories and algorithms based on complex network theory.
However, the integration of D2D and big data
has not been discussed and analyzed
towards a significant volume of real data collected
from practical D2D sharing applications in mobile networks.
Although the online virtual communities of social network services,
in which people with a shared interest, such as a specific hobby or activity, can interact and socialize with each other,
have been researched thoroughly,
the offline realistic people-to-people (i.e., D2D)
content-based interactions can be further socialized
and opportunistically connected toward a promising future
with the assistance of big data techniques.

Therefore, the integration of wireless D2D communications
with big data techniques can extend the limited applications in practice, leading to a new paradigm of offline user-to-user sharing.
And thus, it is necessary to deploy a platform based on big data techniques
(e.g., Hadoop, Spark, MLlib etc.)
for carrying out comprehensive measurement
and analytics based on \textit{large-scale} real-world D2D sharing activities,
in order to enhance the functionality and capability of D2D Big Data
and to improve the quality of D2D services finally.

To the best of our knowledge, we are the first group to study
big data-based large-scale offline D2D content dissemination service in mobile networks.
Our contribution can be summarized as follows:
1) we propose an advanced framework of D2D Big Data platform over cloud-based systems;
2) we discuss potential system and modules for D2D Big Data framework to strongly encourage D2D sharing;
3) we provide valuable measurement, analytics, learning, prediction and recommendation results;
and 4) we discuss useful implications and guidelines for exploiting D2D Big Data for enriching offline services.
In addition, we also discuss opportunities and challenges
to hopefully unveil the upcoming future of wireless D2D communications.
g coverage percentage.

\section{D2D Big Data Framework}
\label{sec:d2dbigdata}

The original objective of utilizing D2D has been just to query neighboring peers for obtaining desired content files,
and also to broadcast interesting or urgent information among mobile users.
Extending the scope of existing D2D studies in Delay-tolerant Networks (DTNs),
Opportunistic Networks (ONs) and Mobile Social Networks (MSNs),
which are mostly designed for offloading and sharing based on small number of users,
we propose to design a D2D Big Data framework for large-scale user base.
It is useful in facilitating the functionality of D2D sharing tools to level up towards
an enhanced offline content dissemination service
and a mobile interest exploitation platform,
which can serve a significant amount of mobile users
over a large area (even nation).

\begin{figure}[!hhhhhhhhhht]
\centering
\includegraphics[width=15cm]{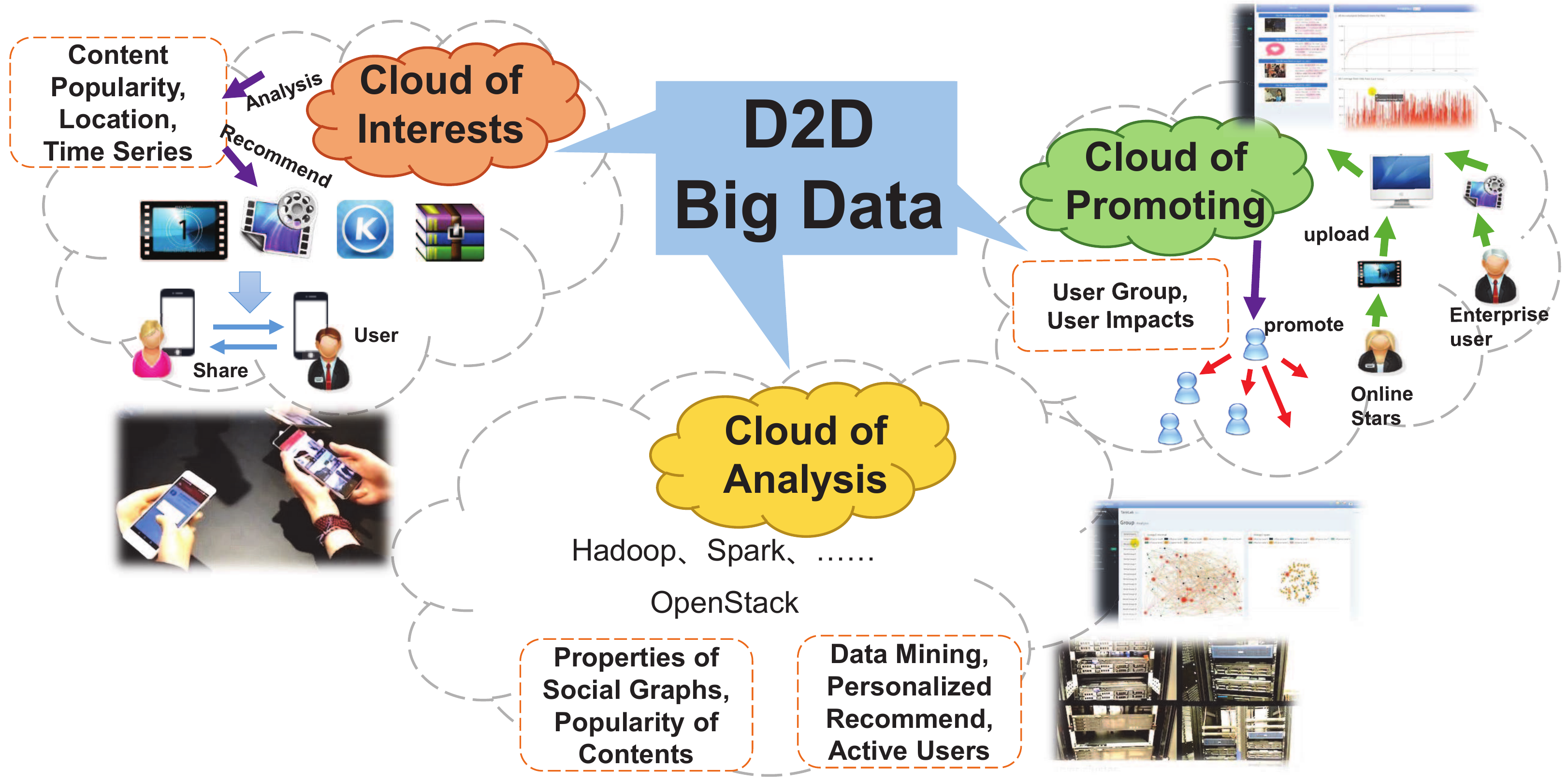}
\caption{Illustration of three essential parts of D2D Big Data framework}
\label{fig_2_threeclouds}
\end{figure}

From the perspectives of mobile users, operators and content providers,
as shown in Fig. \ref{fig_2_threeclouds}, D2D Big Data consists of 3 parts:
\begin{itemize}
\item \textbf{Content Promotion Big Data (\emph{Cloud of Promoting})}:
It is a platform open to customers
who want to promote content effectively and rapidly.
The main functions are content uploading, budgeting, promotion supervising and strategy visualizing.
The target is to help the content provider make better promoting strategies and popularize content by D2D deliveries
in a short time.
Additionally, multidimensional features involving users' interests, relationship,
user impacts, social groups will be analyzed to generate
a proper strategy for content providers to make their promotion decisions.
\item \textbf{Interest Exploitation Big Data (\emph{Cloud of Interests})}:
There is an interest mining and analyzing platform
that focuses on mining users' preferences to realize the functions of interest recommendation, discovery the interest of nearby users, establishing groups and obtaining interesting contents.
The interest exploitation functions are realized by extracting features of users'
interactive traces, their preferences, the sharing habits and content popularity to perform customized and accurate recommendation.
\item \textbf{D2D Analysis Big Data (\emph{Cloud of Analysis})}:
As the core of the scheme, the D2D Big Data analysis platform is
built to process and analyze any data sets and to gain meaningful results that the content providers need.
The main part is a series of algorithms and frameworks of measuring and extracting multi-features from the traces,
including time-varying online behaviors, location relations and similarity, content properties and users' preference entropy,
characteristics of social groups, meeting dynamics, friendship tree analysis etc.
In addition, they are implemented by the Spark-based big data framework to ensure the high performance and accurate prediction.

\end{itemize}

The industrial vision of D2D Big Data is to fix the problems of overloaded
online Internet content,
which have been severely confusing users' demands
and wasting large amounts of marketing costs with low efficiency and precision instead.
Hence to offload valuable content into offline ``face-to-face'' (i.e., D2D) deliveries
due to the truth of word of mouth effect \cite{WordOfMouth_IMC11},
are much more accurate for satisfying mobile users.
And from the academic perspective on D2D Big Data,
the objective is firstly to effectively process large scale data set generated by millions of users;
secondly, it is not only to efficiently harmonize modules of machine learning,
graph computing, complex network theory etc.,
but also to further invent novel theories and techniques.
Therefore, it is desired to deploy the D2D Big Data framework carefully
for satisfying the needs, and we will elaborate details in following sections.

\section{Measurement and Analytics over Realistic Data and System}
\label{sec:measurementandanalytics}

\begin{figure}[!hhhhhhhhhht]
\centering
\includegraphics[width=16cm]{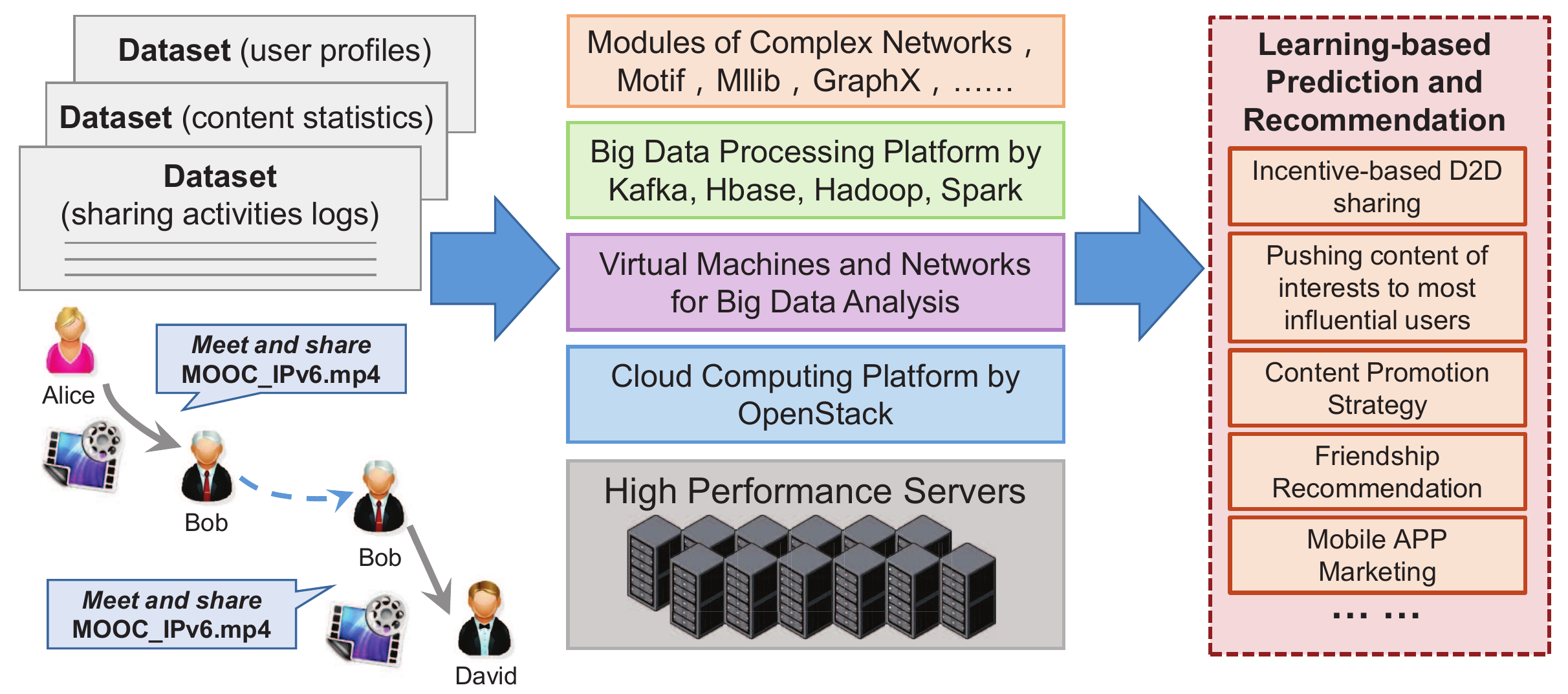}
\caption{The framework of the D2D Big Data system, and the related processing work flow}
\label{fig_3_system}
\end{figure}

The main architecture of D2D Big Data framework is illustrated in Fig. \ref{fig_3_system}.
There are two essential conceptual components, the input of large scale \textbf{data set}
including user profiles, content statistics, activity logs, traffic records etc.
While users move and share content files,
the \textbf{system} carries out streaming-based and batch-based processing.
Prediction and recommendation are made by learning-based tools involving machine learning, unsupervised learning and collaborative group recommendation algorithms
to realize content promotion and friendship recommendation. Meanwhile, analytics and findings can further contribute to the mobile APP marketing and developing strategies.

A robust data processing platform
based on our server clusters and Spark system is built.
To solve the storage problem of large  data sets and processed results,
Hadoop Distributed File System (HDFS) is implemented to ensure the scalability and robustness of the system.
Furthermore, the in-memory computing platform of Spark is deployed to improve the big data processing ability of the framework.
We used Python scripts along with Anaconda scientific package and graph-tool library (v2.18),
Hadoop, Spark, MLlib, GraphX, Spark Streaming and batch-based data processing tools like Kafka,
to support the analytics and computing of the large data.
All tasks are supported by a cluster of 24 servers,
each of which has two 2620v4 CPUs, 64G or 96G RAM and 6T SAS hard disk.

However, we do not deploy big data system on the servers directly,
but firstly implement OpenStack over the servers, and then D2D Big Data processing tools and software packages can be set up on the virtual machines.
We use OpenStack for a cloud computing infrastructure, because it is an open source cloud computing software with functions of
building up the cloud operating system that cloud control large pools of compute, storage, and networking resources throughout a datacenter in the IaaS layer. The virtual machines with the widely used big data processing softwares including Spark, Hadoop and Hive, thus can be elastically created and used through OpenStack.
This way, the platform can enjoy the elasticity and the optimality
of the consolidated cloud resource all the time.

For testing and evaluating D2D Big Data framework,
we capture and import a large scale data set from \emph{Xender},
which is a content delivery mobile application,
serving millions of users all over the world
to share content files via smart phones with ease.
The service has been increasingly supported more than 100 million monthly active users
and approximately 110 million content delivery activities per day.
We measured the data sets from 01/08/2016 to 29/10/2016,
13 weeks in total, and a data set of approximate total size of 3.56 TBytes.
We claim that the statistical information of the service and trace in this article
can be only used for research purposes as it does not exactly reflect a realistic business scale,
because while capturing the trace, we already filtered out
a large number of incomplete or invalid records.
More details of the data set and system are shown in Table \ref{table_1}.
Note that the scale of the data set and system in our experimental lab
are actually not so large compared with industrial big data systems run by big companies.
But we believe they are enough to ensure high computing performance,
memory capacity and robust performance for digging  potential capacity of D2D.

\begin{table}[!hhhhhhhhhht]
\centering
\caption{Details of the Social D2D Big Data Trace and Platform}
\label{table_1}
\begin{tabular}{|l|l|}
\hline
\multicolumn{2}{|l|}{\textbf{Dataset Details}}  \\ \hline
Data Periods             	& 2016.08.01 - 2016.10.29 \\ \hline
Raw Dataset Size    		& Around 3.56 TBytes \\ \hline
Cleaned Dataset Size   	& Around 96 GBytes  \\ \hline
Total Num. of Users		& 854 millions  \\ \hline
Num. of Involved Users  	& 24 millions  \\ \hline
Total Num. of Activities    	& 8.24 billions  \\ \hline
Num. of Involved Activities & 886 millions  \\ \hline
Total Num. of Files       	& 6.22 billions \\ \hline
Num. of Involved Files    	& 4.45 millions  \\ \hline
Num. of GPS Records  	& 4.19 millions  \\ \hline
\hline
\multicolumn{2}{|l|}{\textbf{Platform Details}} \\ \hline
Server Nodes		        	& 24              \\ \hline
CPU model           		& E2620v4            \\ \hline
Core Frequency      		& 2.1 Ghz           \\ \hline
Memory per Node     	& 96 G        \\ \hline
Storage per Node    		& 6 TB     \\ \hline
\end{tabular}
\label{table_datasetplatform}
\end{table}

Multi-dimensional features, including online behaviors of users, location relations,
meeting dynamics, content properties, social characteristics, popularity trends and so on,
are extracted and mined by the \emph{Cloud of Analysis} to exploit the properties of social graphs,
user behaviors and relationships.
User preferences and content properties are then analyzed by the
\emph{Cloud of Interests} and \emph{Cloud of Promoting}, respectively.
We can step to advanced utilization and exploitation over the potential of offline D2D deliveries.
For example, incentive-based promotion, push over most influential users,
interest exploration and so on
are mostly yet not discovered well.

\section{Results of Measurement, Analysis, and Prediction by D2D Big Data Platform}
\label{sec:predictionandpropagation}

Over the proposed D2D Big Data platform,
there have been running out more and more results regarding
general statistical measurement, analytics on users and contents.
We explore the social graph properties by complex network theory,
temporal and spatial user relationships and so on \cite{d2dCCPE}. 
Due to limited space in this article, we cannot show all of our recent results
but several interesting research items regarding predictive content propagation
based on D2D Big Data.

\begin{figure}[!hhhhhhhhhht]
\centering
\includegraphics[width=16cm]{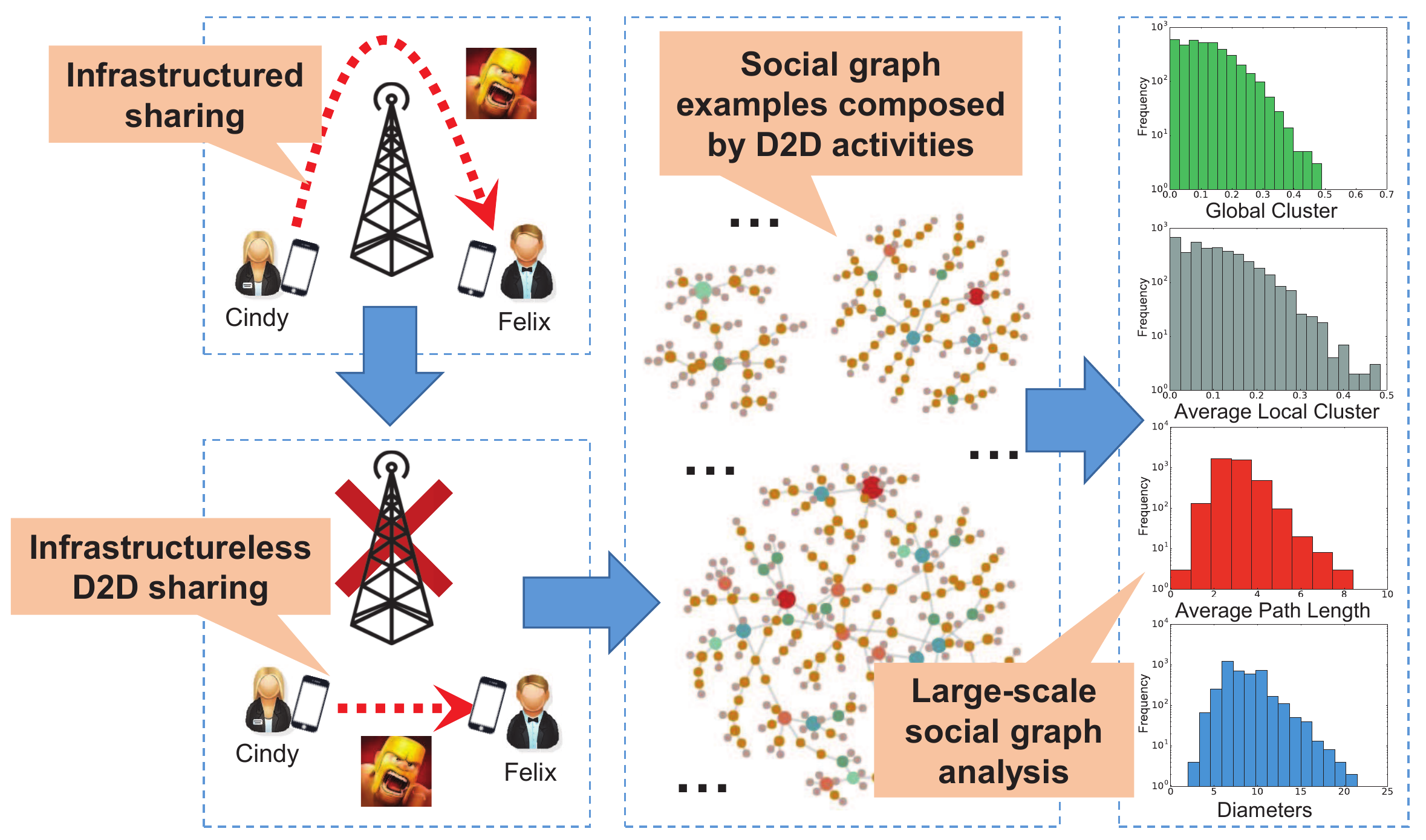}
\caption{Example of social graph composed by D2D activities from infrastructured sharing to D2D sharing.}
\label{fig_4_socialgraph_example.eps}
\end{figure}

\begin{figure}[!!!!!!!!!!!!!!hhhhhhhhhht]
\centering
\subfigure[Reduplicate traffic load via D2D sharings]
{\includegraphics[width=5.6cm,height=4.2cm]{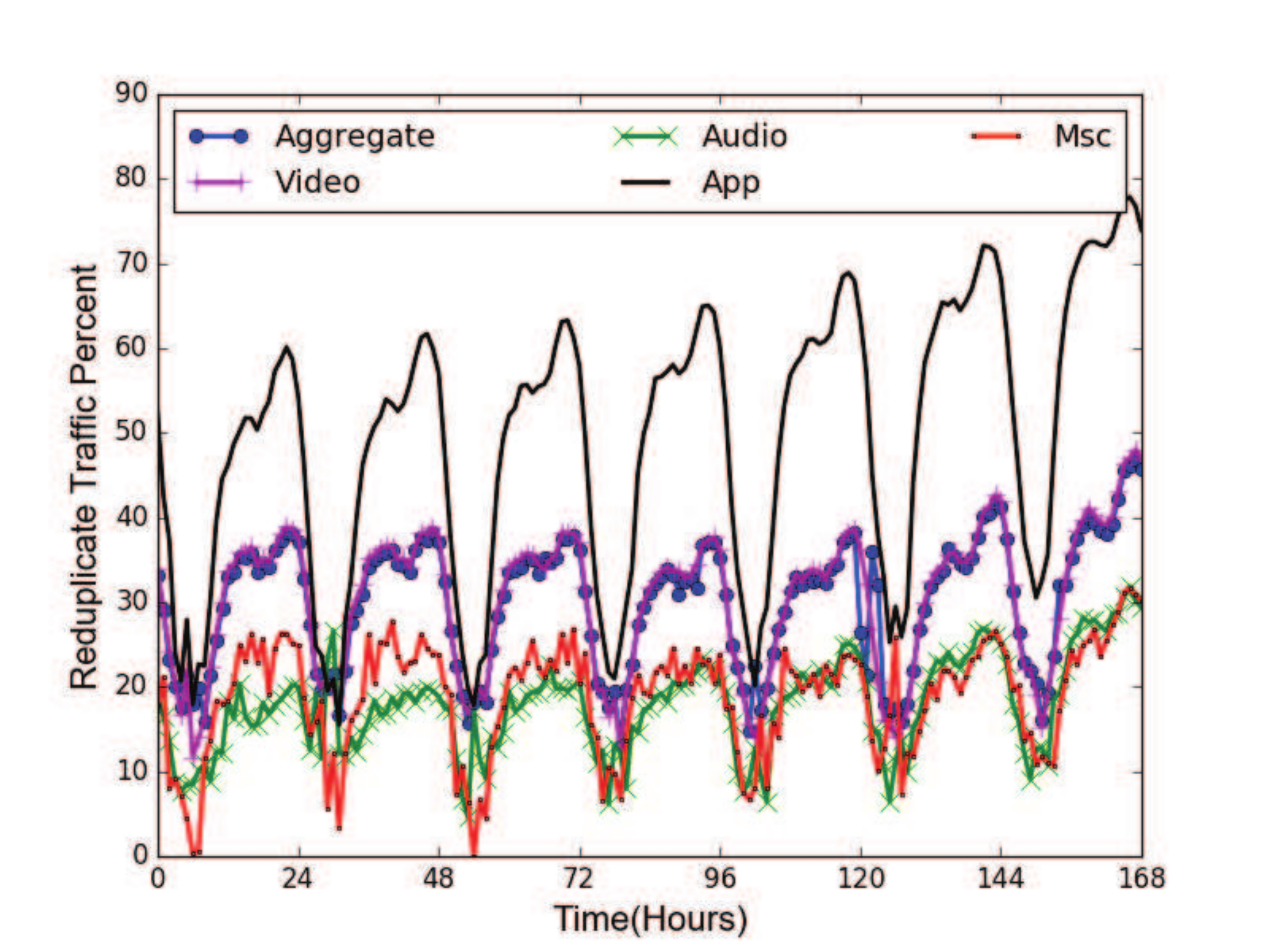}
\label{fig_6_timeseries_duplicatedtrafficload.eps}
}
\subfigure[Groups sizes by vertex]
{\includegraphics[width=5.5cm,height=3.9cm]{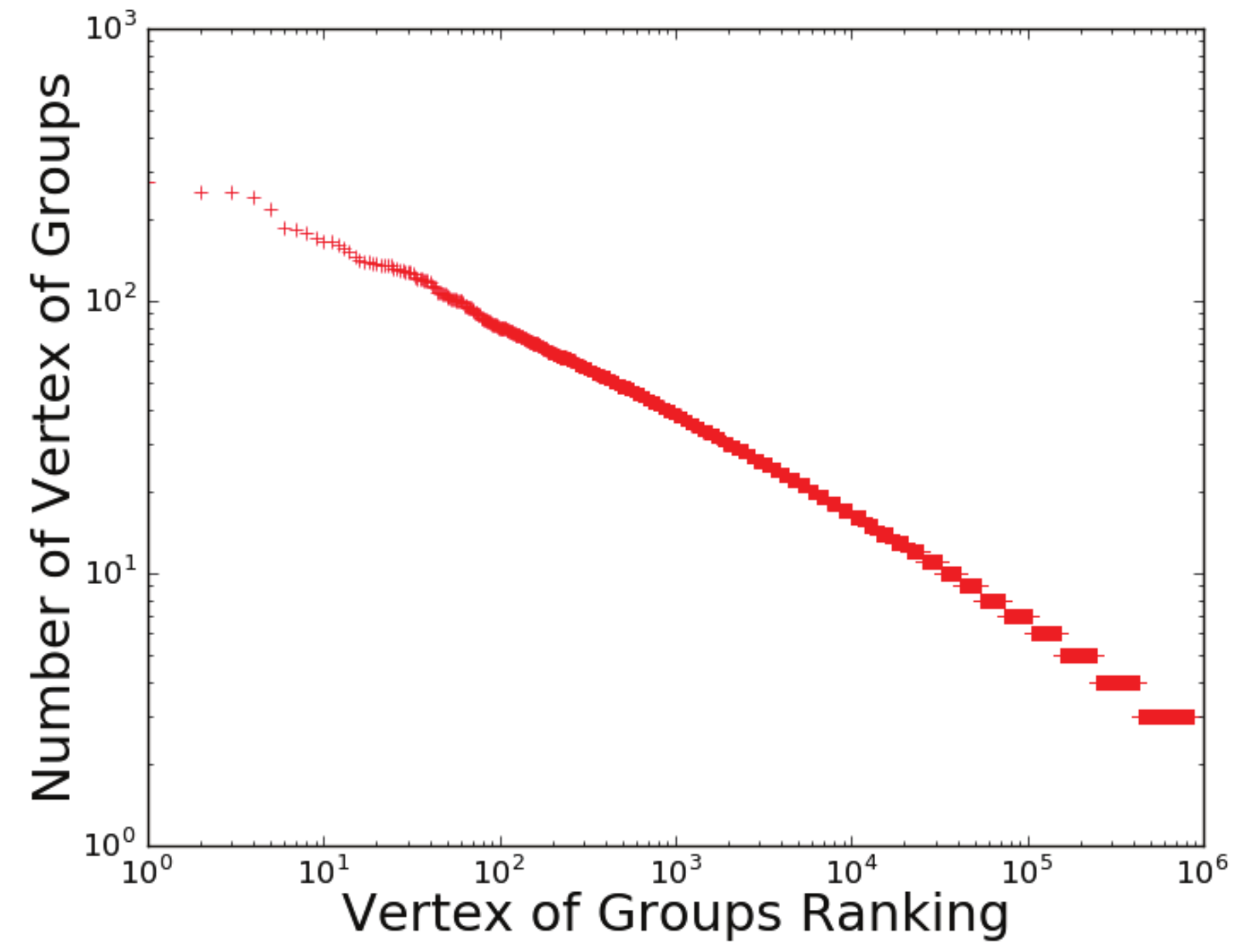}
\label{fig_6_groupsbyVertex.eps}
}
\subfigure[Coverage percentages by selecting effective seeding users]
{\includegraphics[width=6.0cm,height=4.1cm]{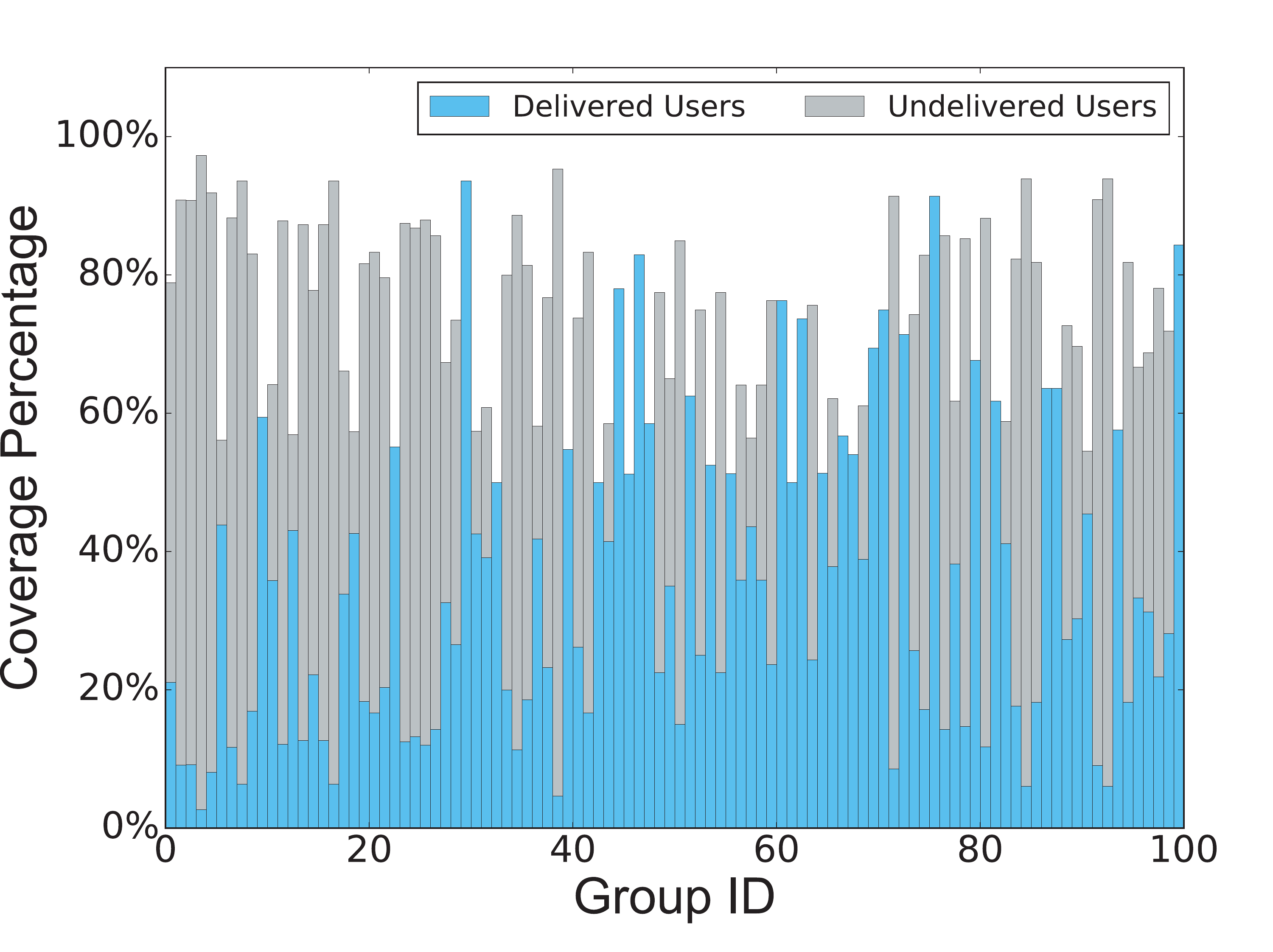}
\label{fig_6_coverage.eps}
}
\caption{Selected results from D2D Big Data platform on measurement, social graph analysis, predictive propagation.}
\label{fig_5_tree}
\end{figure}

From Fig. \ref{fig_4_socialgraph_example.eps}, we can see that
a conventional cellular-based sharing activity can be changed
by wireless D2D communication where the BS is not necessarily needed.
When many mobile users share files via D2D communications,
as time goes on, the wireless links among users formulate ``virtual'' connectivities
that transform the sharing records into an offline social network,
which user behavior pattern and social properties of offline MSNs are quite similar
to online social networks. Due to the temporal and spatial constraints
of D2D deliveries, which are much difficult to carry out compared to online sharing activities,
the graph in the D2D social network is not strongly connected but a bit sparse,
and our strategy is to split just by the physical encounters,
i.e., if two users have ever shared at least one content file via D2D,
they are in the same group.
And thus we can approximately split the large user base into multiple groups,
in which members are tightly connected.
As shown in Fig. \ref{fig_6_groupsbyVertex.eps},
the sizes of the groups follow the power law exactly,
which means there are few groups with large sizes, but most of the groups have small sizes.
This provides us with the space of emphasizing on some system optimization as well as group recommendation policies.
Also we are further applying in-depth social network analysis based on complex network theory,
and try to find out the properties of user groups, for example,
global cluster and local cluster, average  path length and diameters,
which are illustrated inside Fig. \ref{fig_4_socialgraph_example.eps} as well.
For instance, the diameters of those social groups in D2D sharing
are around 6 to 10, which are quite larger than online social network services (SNSs),
indicating that the sharing by D2D happens not as easy as that in online SNSs.
Instead, D2D activities take place with strict conditions on space and time,
as well as very strong demands among users.

\begin{figure}[!hhhhhhhhhht]
\centering
\includegraphics[width=18cm]{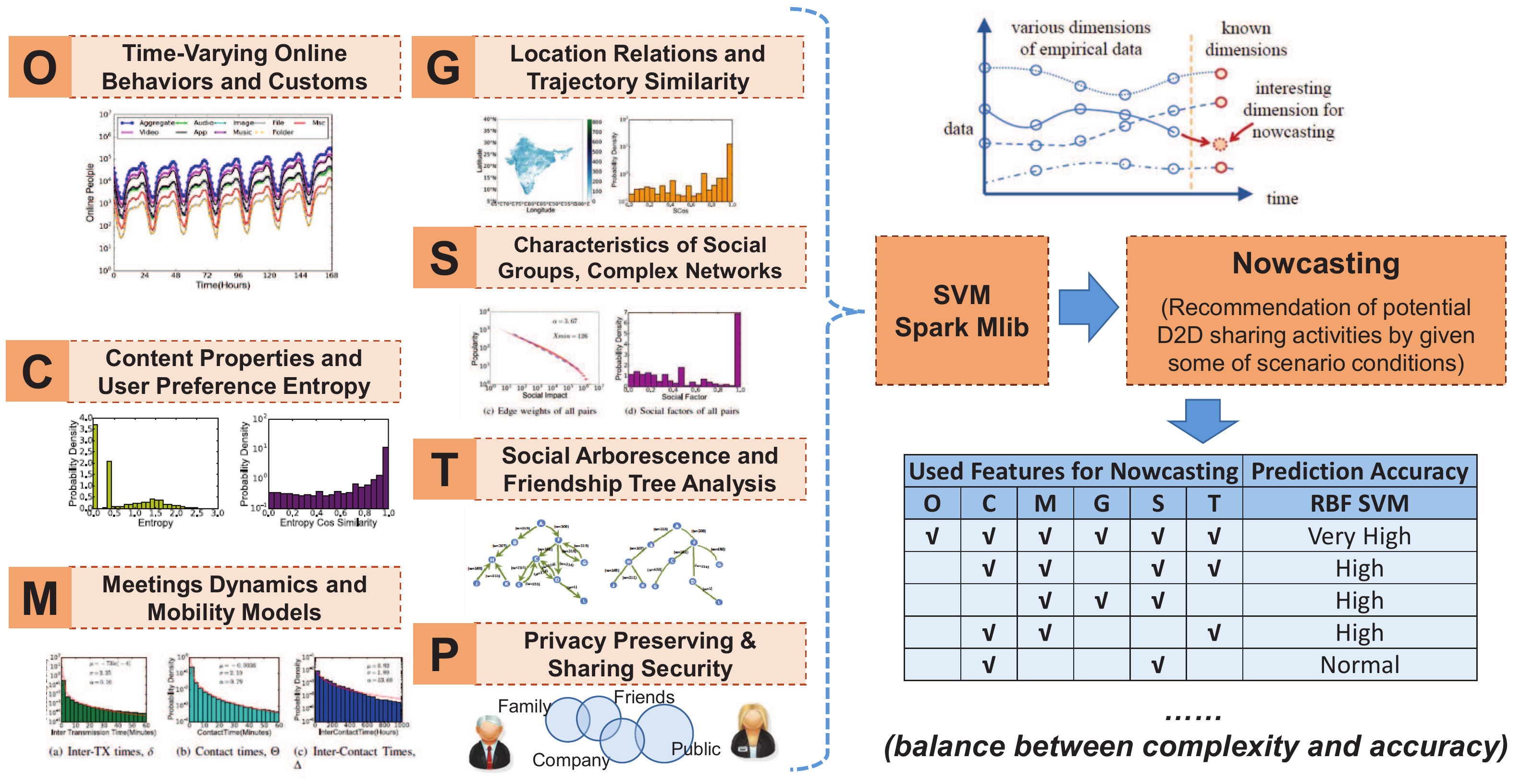}
\caption{Procedure of multi-feature learning and prediction of wireless social D2D deliveries based on D2D Big Data.}
\label{fig_5_framework_2}
\end{figure}

The key function driven by D2D Big Data must be the accurate prediction of user activities.
This way, platform resource can be allocated in prior to the practically happening,
and also we can show proper recommendation system to improve the quality of service.
We consider several important features, including time-varying online behaviors and customs,
content properties and user preference entropy,
meeting dynamics and mobility models,
location relations and trajectory similarity,
characteristics of social groups by complex networks,
social arborescence and friendship tree analysis as well as privacy preserving and sharing security as shown in Fig. \ref{fig_5_framework_2}.
The privacy preserving and sharing security here can be realized by setting different permission levels
to users according to their relationships.
Only users with the same permission levels like family or friends can access to others' content files and hence the privacy is ensured.
We are trying to carry out nowcasting experiments on D2D sharing activities,
based on SVM and MLlib, by given several happened factors \cite{d2dCCPE}.
As shown  in Fig. \ref{fig_5_framework_2}, multiple features are extracted from the empirical data (e.g. 6 weeks' data) to train the learning tools of SVM or Spark Mlib.
And we nowcast the potential D2D sharing activities among users with test data (e.g. 7 weeks' data).
For example, when a user who likes playing free mobile games is taking a rest at a cafeteria,
we can analyze and predict a potential D2D sharing of the user
with nearby users with common interests,
and thus utilize push service for encouraging an D2D sharing activity with highest probability.
Results in Fig. \ref{fig_5_framework_2} show that with different combinations of the features,
the prediction accuracy varies.
From the perspective of efficiency,
when we desire higher prediction rate of forecasting by multiple features,
we can selectively analyze 2 or 3 features mentioned above, not all of them.
There is a significant tradeoff between features put into the model.

We can easily obtain a number of statistical measurement results time by time.
One interesting finding is that,
as shown in Fig. \ref{fig_6_timeseries_duplicatedtrafficload.eps},
there is huge amount reduplicate transmissions among millions of users.
Particularly APPs and videos have very high percentage of reduplicate
traffic, which indicates the skewed content popularity of the two types,
It may motivate us to apply related prioritization algorithms for related resource optimization,
like CDN deployment and content recommendation strategies.

Another key function of D2D Big Data is
to find the most influential users within a large number of mobile users,
considering various properties of the users, content files, groups, and so on.
We have designed algorithms to discover the seeding users \cite{d2dCCPE}
by finding the main roots of the tree structures in the social groups of the large user base
from the first half of the trace.
Fig. \ref{fig_6_coverage.eps} is the results of the evaluation
on the efficiency of propagating contents
in 100 randomly selected groups as representatives,
and note that we only find the most influencing seed for each group
to spread the content to others,
by assuming that two users may share the content once they practically encounter in the second half of the trace.
It is clearly shown that in the groups,
nearly a half of the mobile users can get the content via offline D2D sharings,
and it appears similar patterns of spreading popular contents rapidly boosted by
seeding users' recommendations in online SNSs \cite{RecommdBoostSPSNS_WWW12}.

\section{Opportunities and Challenges}
\label{sec:opportunitiesandchallenges}

We hereby discuss a few essential promising research directions of D2D Big Data,
and highlight several pending problems from the perspective of elaborating and widening the usage of D2D techniques.

\subsection{Privacy, Security and Energy Issues for D2D Big Data}
Despite the tremendous benefits of D2D deliveries and the attached values of Big Data-driven intelligence,
D2D still has many security and privacy challenges that are not yet well solved.
There have been many studies on privacy and security issues over DTNs/ONs/MSNs,
like \cite{security_more} and \cite{rlu_rrefilter_PrivacyDTN_infocom12}.
Nonetheless, the problems may be enlarged seriously when the platform grows to a huge one
for nation-wide or even world-wide MNOs,
big content provider companies, and millions of mobile users.
Offloading content traffic onto D2D with big data techniques
should provide a fair mechanism for appropriately negotiating
the cost, profit, and quality of service among all the involved entities.
Particularly, the effective balance between the privacy protection and freedom of sharing
in mobile devices should be dug harder,
and also the tradeoff between passive transparent sharing or pull-based active sharing
via D2D should be discussed with more experiments and user-centric trials.
The achievement of a ``triple-win'' for the safe and secure
D2D industry benefiting content providers, users and the platform developers will be complicated, and the success highly relies on
innovative privacy schemes, copyright mechanism, and security protections for D2D Big Data for a large-scale user base,
which are still undiscovered.
One more critical issue of D2D sharing is the precious
energy of mobile devices \cite{D2denergyyzhang},
and hence how to utilize the emerging energy harvesting techniques
for D2D communications becomes interesting but challenging.

\subsection{Big Data-assisted Base Station Caching Integrated with D2D}
There have been a number of studies towards
the Mobile Edge Computing, Content-Centric Networks and so on,
which mostly highly rely on the caching and offloading
capability on the base stations (BSs) \cite{cacheintheair}.
BSs' edge cache can formulate a collaborative ``buffer'' to significantly optimize the mobile applications by the last hop.
There have been a few proposed techniques and systems
for integrating edge caching functionality into MNO's network to analyze and facilitate the mobile services;
however, it is still not discussed or even tested to combining BS caching with D2D communications
together for a more pervasive and more universal caching space around mobile users.
We are faced with interesting but challenging research issues,
like how to optimally allocate the networking and caching resource among BSs and smart devices,
and how to even deploy big data techniques
to analyze and disseminate contents to a nation-wide coverage of mobile users with the best performance.

\subsection{Business Models for D2D Big Data}
There has been an explosive growth of the online peer-to-peer (P2P)
techniques since the beginning of 21th centaury,
along with the growth of Internet topology and the number of Internet hosts,
which has enabled people to collaborate
for sharing contents online massively in a distributed manner.
For now, can we expect a second round of P2P,
the thriving offline D2D deliveries via face-to-face encounters among people?
One motivation for mobile users to share content with each other
is the common-interest, but beyond that D2D Big Data
should carry out fine-grained incentive-based forwarding mechanism
to push the content promotion activities naturally.
For example, while there are so many mobile application (APP) markets online,
but can we distribute most of the free APP files (i.e., APK files for Android)
via user-to-user sharing \cite{thanasis_mobileapp_ecosystem}.
Also can we deliver music and video files to proper audiences,
while each user is both a viewer and a marketing participant?
More business models over D2D Big Data should be explored in the near future.

\section{Conclusion}
\label{sec:sonclusion}
In this article, we discussed the potential of
utilizing wireless Device-to-device (D2D) communications
to offload mobile traffic and to disseminate contents
among mobile users assisted by the integration of big data techniques.
We discussed the D2D Big Data framework,
which can encourage the D2D deliveries among users effectively,
promote contents for providers accurately,
and carry out offloading intelligence for operators efficiently.
We implemented a realistic experimental D2D Big Data platform,
and exploited a significant volume of real data set.
Based on the measurement, analysis and evaluation,
we exploited and discussed important implications,
under-going techniques, opportunities and challenges
to improve quality of D2D sharing services,
and to enhance the functionality and capability of D2D Big Data.
For our future work, we concentrate on
not only more satisfactory group recommendation techniques
and more accurate prediction of content popularity and user behaviors,
but also the consolidated multi-disciplinary collaboration
of D2D Big Data with MNOs' wireless systems,
industrial vendors, content service providers, and governments' copyright policies.

\section*{Acknowledgement}
This work was supported by National Natural Science Foundation of China (No. 61702364)


\begin{biography}
Xiaofei Wang [M'13] (xiaofeiwang@tju.edu.cn)
is currently a professor in Tianjin University, China.
He received M.S. and Ph.D degrees from the
School of Computer Science and Engineering, Seoul National
University in 2008 and 2013 respectively. He received the
B.S. degree in the Department of Computer Science and Technology, Huazhong University
of Science and Technology in 2005. He is the winner of the
IEEE ComSoc Fred W. Ellersick Prize in 2017.
His current research interests are social-
aware multimedia service in cloud computing, cooperative
backhaul caching and traffic offloading in mobile content-
centric networks.
\end{biography}

\begin{biography}
Yuhua Zhang [S'17] (yuhuazhang@tju.edu.cn)
is a master student in Tianjin University, China.
She got bachelor degree in Shandong Normal University, China.
Her research interest is social device-to-device communications and big data.
\end{biography}

\begin{biography}
Victor C.M. Leung [S'75, M'89, SM'97, F'03] (vleung@ece.ubc.ca) is a
Professor of Electrical and Computer Engineering and holder
of the TELUS Mobility Research Chair at the University of British Columbia (UBC).
His has co-authored more than 900 journal and conference papers in the areas of wireless networks and mobile systems.
Dr. Leung is a Fellow of the Royal Society of Canada,
the Canadian Academy of Engineering and he Engineering Institute of Canada.
He is an editor of IEEE JSAC-SGCN, IEEE Wireless Communications Letters and several other journals.
\end{biography}

\begin{biography}
Nadra Guizani (nguizani@purdue.edu) is a Ph.D. student and
a graduate lecturer in the Electrical and Computer Engineering Department at Purdue University.
Her research work is on data analytics and prediction and access control of disease spread data on dynamic network topologies.
Her interests include machine learning, mobile networking,
large data analysis, and prediction techniques.
She is an active member of the Women in Engineering Program.
\end{biography}

\begin{biography}
Tianpeng Jiang (tianpengjiangxender@gmail)
is the CEO of Beijing Anqi Zhilian Technology Co. Ltd., Beijing, China.
He got master degree in Beihang University, Beijing, China.
Mr. Jiang has rich industrial experience on social device-to-device sharing techniques.
\end{biography}

\end{document}